\documentclass[prb,citeautoscript,twocolumn,groupedaddress]{revtex4-1} 
\usepackage{graphicx}
\usepackage{units}
\usepackage[intlimits]{amsmath}
\usepackage{amssymb}
\usepackage{hyperref}
\usepackage{siunitx}
\usepackage{lineno}
\usepackage{multirow} 
\usepackage{siunitx} 
\usepackage{natbib}
\usepackage[usenames, dvipsnames]{color}
\usepackage{comment}

\usepackage{hyperref}
\usepackage{amsmath}

\begin{document}

\title{Broadband reduction of quantum radiation pressure noise via squeezed light injection}

\author{Min~Jet~Yap$^{1*}$, Jonathan~Cripe$^{2}$, Georgia~L.~Mansell$^{3,4}$, Terry~G.~McRae$^{1}$, Robert~L.~Ward$^{1}$, Bram~J.J.~Slagmolen$^{1}$, Daniel~A.~Shaddock$^{1}$, Paula Heu$^{5}$, David Follman$^{5}$, Garrett D. Cole$^{5,6}$, David~E.~McClelland$^{1}$, and Thomas~Corbitt$^{2}$}

\address{
$^{1}$OzGrav, Department of Quantum Science, Research School of Physics and Engineering, Australian National University, Acton, Australian Capital Territory 2601, Australia\\
$^{2}$Department of Physics \& Astronomy, Louisiana State University, Baton Rouge, LA 70803, USA\\
$^{3}$LIGO Hanford Observatory, P.O. Box 159, Richland, Washington 99352, USA\\
$^{4}$Massachusetts Institute of Technology, Cambridge, Massachusetts 02139, USA\\
$^{5}$Crystalline Mirror Solutions LLC and GmbH, Santa Barbara, CA, 93101 and 1060 Vienna, Austria\\
$^{6}$Vienna Center for Quantum Science and Technology (VCQ), Faculty of Physics, University of Vienna, A-1090 Vienna, Austria\\
$^*$Corresponding author: minjet.yap@anu.edu.au
}

\begin{abstract}

We present the reduction and manipulation of quantum radiation pressure noise (QRPN) in an optomechanical cavity with the injection of squeezed light. The optomechanical system consists of a high-reflectivity single-crystal microresonator which serves as one mirror of a Fabry-Perot cavity. The experiment is performed at room temperature and is QRPN dominated between 10 kHz and 50 kHz, frequencies relevant to gravitational wave observatories. We observed a reduction of 1.2 dB in the measurement noise floor with the injection of amplitude squeezed light generated from a below-threshold degenerate optical parametric oscillator. This experiment is a crucial step in realizing the reduction of QRPN for future interferometric gravitational wave detectors and improving their sensitivity.

\end{abstract}

\maketitle




\section*{Introduction}

	Effects due to quantum mechanics are becoming significantly important in the precision measurement of continuous variables. As the precision of an observable increases, a back action effect governed by the Heisenberg uncertainty principle results in a increased uncertainly in the conjugate variable. This can be observed in optomechanical systems where the mechanical motion of an oscillator is coupled to an optical cavity mode \cite{Optomechanics}, such as gravitational wave (GW) interferometers. Increasing the laser drive power lowers the photon counting uncertainty and reduces shot noise. The increased power, however, results in an increase in the back action effect in the form of quantum radiation pressure noise (QRPN) \cite{Caves_1980, Braginsky_book}.
	
	When GW detectors such as the Advanced Laser Interferometer Gravitational Wave Observatory (LIGO) \cite{LIGO}, Advanced Virgo \cite{VIRGO}, and KAGRA \cite{KAGRA}, reach their design sensitivity, quantum noise will be the dominant noise source across most of the detection band, with QRPN dominating at low frequencies between 10 Hz and 100 Hz. This quantum noise arises from vacuum fluctuations which couple to the interferometer via the dark readout port. The injection of squeezed vacuum into the interferometer dark port allows the quantum noise to be manipulated \cite{Caves1981}. Squeezed injection has been demonstrated to reduce the shot noise level of previous generation of GW detectors at both LIGO Hanford \cite{LIGO_SQZ}, and GEO-600 \cite{GEO_SQZ}, and is currently being implemented in current GW detectors. Other QRPN mitigation techniques such as variational readout \cite{Kimble}, conditional squeezing \cite{Yiqiu}, and the use of negative mass systems \cite{Negative_mass}, have also been proposed to improve the low frequency sensitivity of GW detectors.
	
	As GW detectors approach their design sensitivity, it is important study the effects of QRPN to help decide which QRPN reduction technique to employ. The effects and manipulation of QRPN has only been recently observed on tabletop experiments \cite{Purdy, Teufel, Clark_QRPN_SQZ, Purdy_room_T, Sudhir, Cripe_QRPN} as it is typically dominated by mechanical thermal noise and other classical noise sources such as seismic vibrations. However, many of the previous observations of QRPN were made in high frequencies (MHz-GHz), around the mechanical resonance, and thus are not fully applicable for GW detectors which will be QRPN dominated over a large frequency band away from the mechanical resonance. A measurement of QRPN away from the mechanical resonance of the oscillator and at frequencies in the GW band has only recently been performed \cite{Cripe_QRPN}.
	
	Here, we investigate the injection of squeezed light in a QRPN limited optomechanical system, and report the reduction and manipulation of broadband QRPN away from the mechanical resonance and at frequencies relevant to gravitational wave detectors. Our experiment utilizes low-loss single-crystal microresonators with low structural noise property for the effects of QRPN to be observed at room temperature.

\section*{The experiment} 

\begin{figure*}
	\centerline{\includegraphics[width=100 mm]{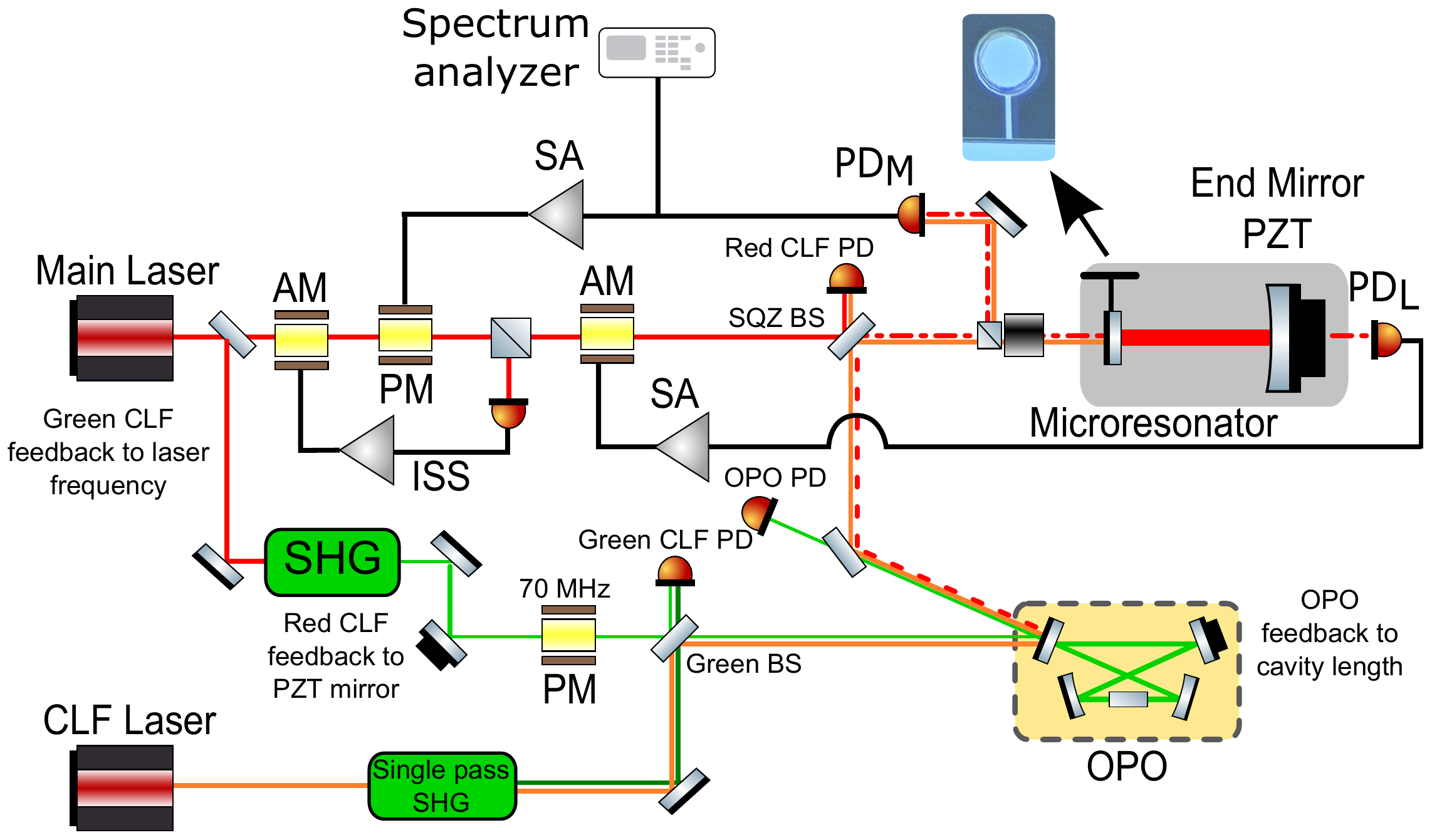}}
	\caption{Schematic of the experiment. The generated squeezed light is produced at the frequency of the main 1064 nm laser (red line). The OPO is pumped by light from the main laser that has been frequency-doubled to 532 nm (light green line) by a SHG cavity. The OPO is locked to the pump field via Pound-Drever-Hall locking with 70 MHz phase modulated (PM) sidebands and the reflected field detected at OPO PD. The generated squeezed vacuum (dashed red) from the OPO is then recombined with the main laser field at a 97:3 beamsplitter to produce a bright squeezed field (dotted dash red) which is then injected into the in-vacuum optomechanical cavity. In order to control the squeezing quadrature of the bright field, a coherent locking scheme was implemented that utilizes a Coherent Locking Field (CLF) laser, frequency shifted by 12.5 MHz from the main laser (orange line), which co-propagates with the squeezed field. The frequency difference between the two laser is maintained by stabilizing the 25 MHz beat note at Green CLF PD. The squeezing quadrature of the bright field is stabilized by controlling the beat note phase of the two lasers detected at Red CLF PD.}
	\label{fig:Setup}
\end{figure*}

	Figure \ref{fig:Setup} shows the schematic of the experiment. The optomechanical system is a Fabry-Perot cavity with a micro-mechanical oscillator as one of the end mirrors. The system is installed on a suspended breadboard inside a vacuum chamber at $10^{-7}$ Torr in order to provide passive seismic and acoustic isolation. The microresonator consist of a roughly 70-$\mu$m diameter mirror pad suspended from a single crystal GaAs cantilever with a thickness of 220-nm, width of 8-$\mu$m, and a length of 55-$\mu$m. The mirror pad is made up of 23 pairs of quarter-wave optical thickness GaAs/Al\textsubscript{0.92}Ga\textsubscript{0.08}As layers for a transmission of T = 250 ppm and exhibits both low optical losses and a high mechanical quality factor \cite{cole08, cole12, cole13, cole14, Singh_PRL}. The microresonator has a mass of 50 ng, a natural mechanical frequency of $\Omega_m = 2\pi \times 876$ Hz, and a measured mechanical quality factor of $Q_m = 16000$ at room temperature \cite{Cripe_QRPN}. The cavity has a length of slightly less than 1 cm, a finesse of $\mathcal{F} = 13000$ and linewidth (HWHM) of $2\pi \times 500$ kHz. 
	
	The optomechanical cavity is operated blue-detuned from a 1064 nm Nd:YAG laser which results in a strong optical spring effect. The optical spring self-locks the cavity for frequencies below the optical spring resonance, however the phase lag due to the finite cavity response results in a anti-damping force, rendering the system unstable \cite{13,17}. The optical spring effect is stabilized by monitoring the cavity reflection and transmission field, and providing active feedback around the optical spring frequency to the laser power and frequency via an electro-optic amplitude modulator (AM) and phase modulator (PM) \cite{Cripe_RPL, Cripe_QRPN}. In the final measurement configuration, only the reflected light and PM feedback loop is used to lock the cavity at a detuning of about 0.6 linewidths, with the optical spring pushing the mechanical resonance frequency above 100 kHz.
	
	The squeezed vacuum state is generated from a sub-threshold degenerate optical parametric oscillator (OPO) via the parametric down conversion process. The OPO is a doubly resonant bow tie cavity with a nonlinear crystal made of periodically poled potassium titanyl phosphate (PPKTP) embedded within the cavity. The OPO is pumped by light tapped from the main laser that has been frequency-doubled to 532 nm via a second harmonic generation (SHG) cavity, and is kept on resonance with the pump light via a Pound-Drever-Hall locking scheme \cite{Drever1983}. Squeezed light is injected into the cavity by combining the main laser field with a squeezed vacuum state via an asymmetric 97:3 beamsplitter. Spatial mode mismatch is filtered out by passing the combined field through a short optical fiber before the optomechanical cavity. An intensity stabilization servo (ISS) is used to suppress the main laser intensity noise down below shot noise level. 
	
	The control of the squeezed ellipse phase with respect to the main laser is achieved with a coherent locking scheme \cite{Chua11, Henning} which utilizes a coherent locking field (CLF) laser frequency shifted from the main laser by 12.5 MHz. The frequency difference between the two lasers is maintained by up-converting a small portion of the CLF laser to 532 nm and phase locking the 25 MHz beat note between the up-converted field and the OPO pump field. The unconverted (1064nm) CLF beam co-propagates with the squeezed vacuum field and is phase locked with the main laser after the asymmetric beamsplitter at 12.5 MHz. Engaging both the CLF phase locks allows the squeezed ellipse to the track the phase of the main laser field. Rotation of the squeezed ellipse between the amplitude and phase quadrature is achieved by changing the demodulation phase between the two CLF phase locks.

\section*{Results}


\begin{figure}
	\centerline{\includegraphics[width=80 mm]{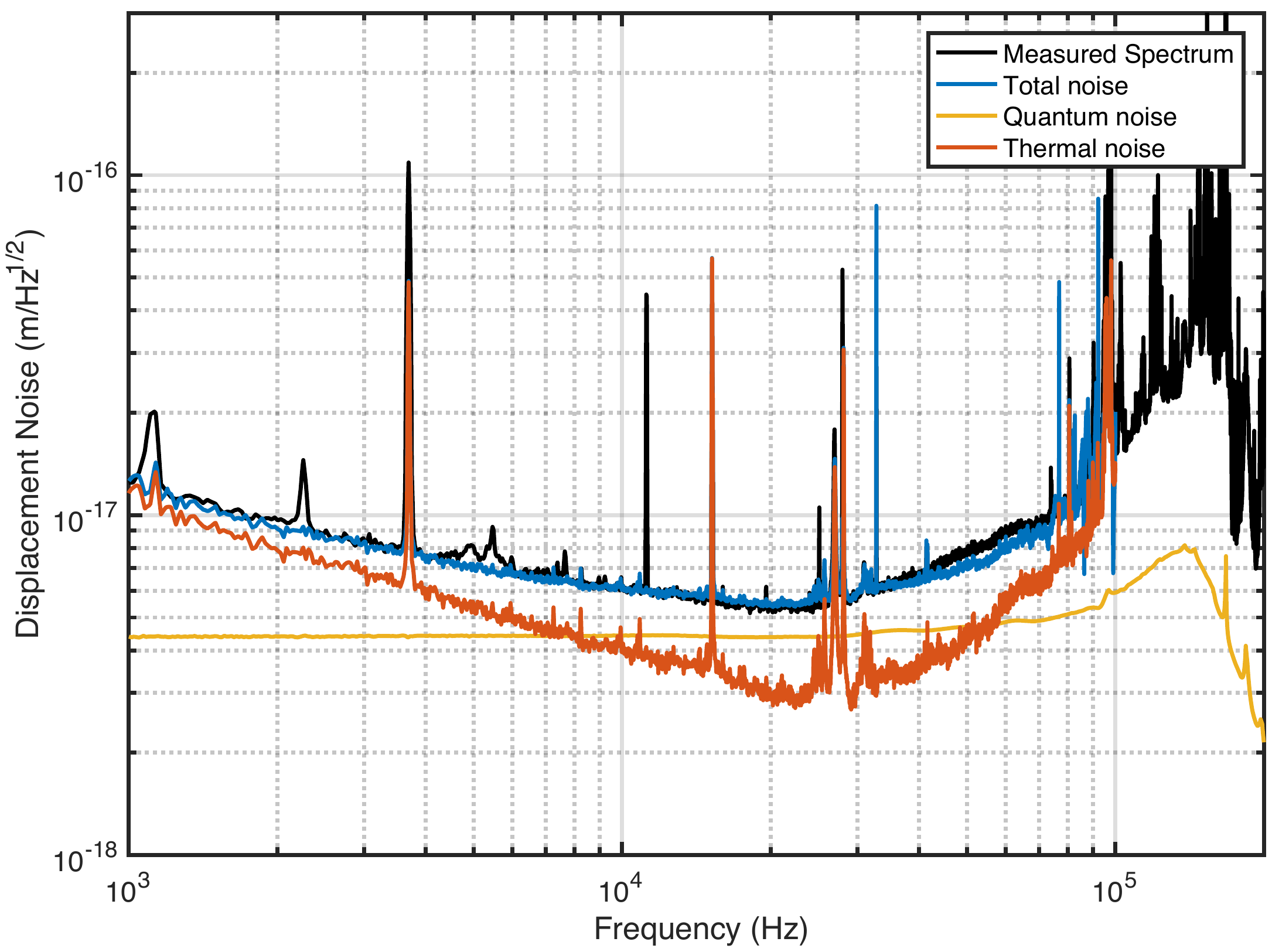}}
	\caption{Displacement spectral density of the microresonator with a 150 kHz optical spring resonance. The dominant noise sources are thermal noise (orange trace), and quantum noise (yellow trace) which is dominated by QRPN below the optical spring resonance. The thermal noise measurement was taken with low intracativity power, and closely follows a structural damping model\cite{Cripe_QRPN}. The quantum noise trace also takes into account the dark noise level of the photodetector. The quadrature sum of the two noise sources (blue trace) is overlaid with the measured displacement spectrum (black trace) with no squeezed light injection.} 
	\label{fig:budget}
\end{figure}

\begin{figure}
	\centerline{\includegraphics[width=80 mm]{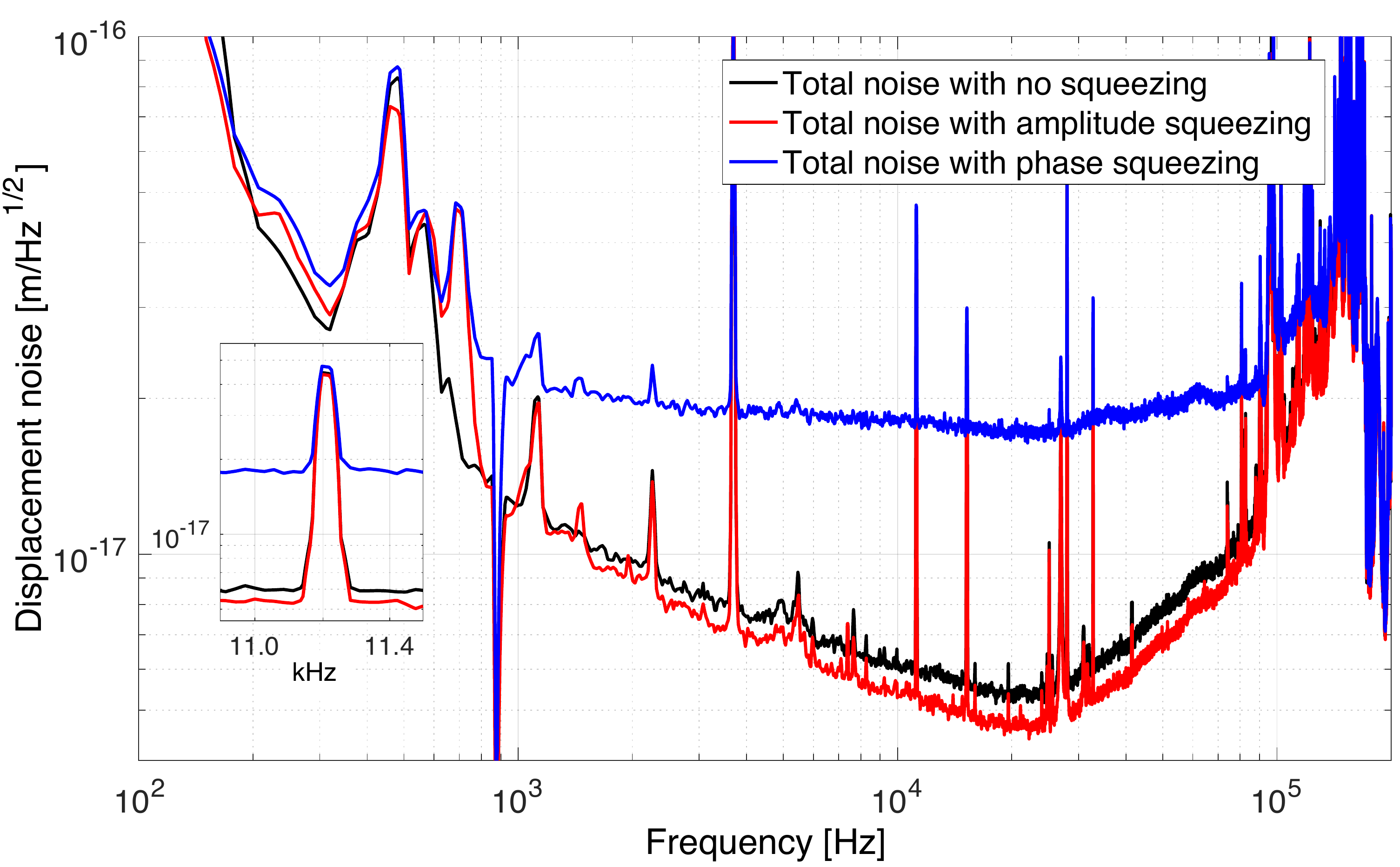}}
	\caption{Displacement noise spectrum of the microresonator with the injection of squeezed light. Reference trace (black) is measured by tuning the OPO to its anti-resonance to ensure no squeezed light is generated. Injection of amplitude squeezed light (red) results in a maximum noise reduction of 1.2 dB at 20 kHz. Rotating the squeezed ellipse to the phase quadrature (blue) increased the total noise of the system. Inset: 11.2 kHz calibration line used in the three measurements.} 
	\label{fig:QRPNtrace}
\end{figure}

	Figure \ref{fig:budget} shows the displacement spectral density measured at the reflection port of the cavity with 220 mW of  circulating power. The broad peak at 150 kHz is due to the mechanical fundamental frequency being shifted up by the optical spring effect \cite{Cripe_RPL}. The dominant noise source below 10 kHz is the thermal noise of the microresonator which follows a structural damping model between 200 Hz to 30 kHz, and falls off as $1/f^{1/2}$ compared to QRPN \cite{Cripe_QRPN}. With 220 mW of circulating power, QRPN is dominant noise source between 10 kHz and 50 kHz. The excess thermal noise above 30 kHz is believed to be related to thermoelastic damping.
	
	The spectrum is calibrated by measuring the transfer function from the main laser piezo to the cavity reflection port. The laser piezo actuates on the main laser frequency and has been calibrated separately. The transfer function measures the closed loop response of the system, and undoes the effect of both the electronic feedback and the optical spring response. The optical spring effect is reintroduced in the spectrum by measuring separately the optical spring frequency and cavity detuning. A 11.2 kHz dither tone on the cavity length is used to produce a calibration line, shown in the inset of Figure \ref{fig:QRPNtrace}, to ensure the calibration is constant between all the measurements.

\begin{figure}
	\centerline{\includegraphics[width=85 mm]{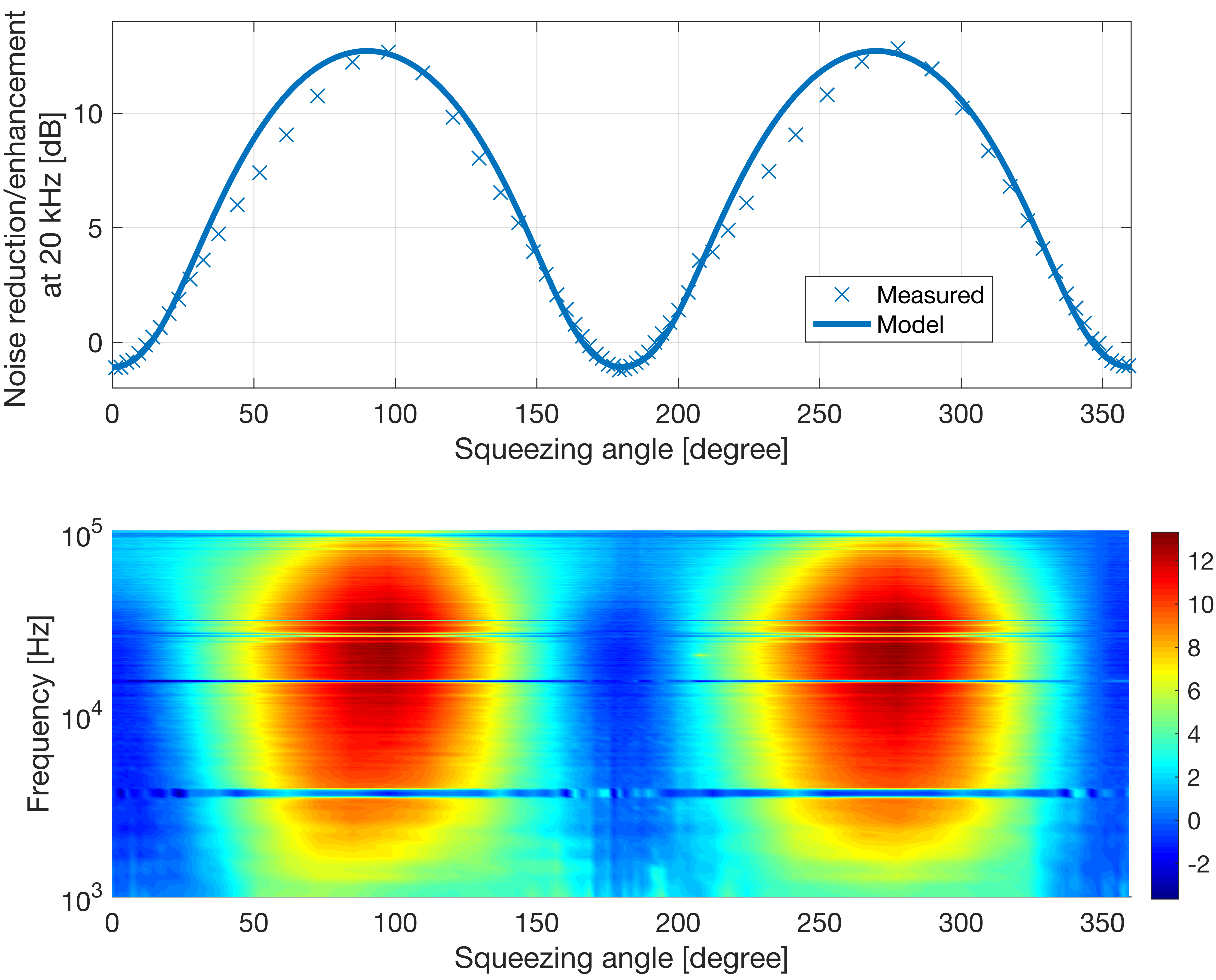}}
	\caption{Displacement noise spectrum as a function the squeezing phase normalized to the reference spectrum at 20 kHz and across the measurement frequency band. Horizontals lines unchanged with the squeezing phase corresponds to the higher order mechanical modes of the microresonator.}
	\label{fig:Arches}
\end{figure}

	
	In order to manipulate the QRPN, bright squeezed light is injected into the cavity, which affects the measured displacement spectrum as shown in Figure \ref{fig:QRPNtrace}. With the injection of amplitude squeezed light, we observe a reduction of the total noise floor at frequencies where QRPN is dominant, with a maximum reduction of 1.2 dB at around 20 kHz. Even though thermal noise is the dominant noise source below 10 kHz, QRPN is still a major contributor to the total noise and the reduction in noise due to squeezed light injection remains visible below 2 kHz. By changing the relative phase between the two CLF locks, we are able to rotate the squeezing ellipse to produce phase squeezed light resulting in an increase of the total noise by 12.6 dB at 20 kHz. The flat and broadband nature of the increase is indicative of the quantum noise being manipulated. Figure \ref{fig:Arches} shows the noise reduction and enhancement  at 20 kHz and across the measurement spectrum.
	
	The amount of observed reduction in noise is currently limited by the collective losses of the system, which degrades the squeezed state by mixing it with uncorrelated vacuum fields. These losses include the OPO cavity escape efficiency, optical propagation loss from the OPO to the photodetector, mode matching efficiency of the squeezed field to the optomechanical cavity, and the photodiode quantum efficiency. The optical propagation loss was measured to be 47\%, which was predominately due to optical fiber launching efficiency, and diffraction losses at the cantilever mirror. The OPO escape efficiency, a measure of the the OPO out-coupling efficiency had a measured value of 97\%, mode matching efficiency to the optomechanical cavity was 80\%, and the photodiode quantum efficiency was 97\%. This resulted in a total loss efficiency of 40\%, which is in agreement with the measured amplitude and phase squeezing level.

\section*{Conclusion}

	We present the reduction of quantum radiation pressure noise of a microresonator far from the mechanical resonance frequency over a broad frequency range via the injection of squeezed light. This provides useful insight in reducing the radiation pressure forces of future gravitational wave observatories in order to improve its sensitivity and detection range. Moreover, a radiation pressure noise limited optomechanical system provides a useful testbed for other QRPN reduction proposals \cite{Braginsky547, Braginsky_speedmeter, Kimble, Harms, FD_SQZ, Glasgow_speedmeter} and quantum-enhanced displacement sensing \cite{Giovannetti1330}.	
	
	With the optomechanical system at room temperature, the standard quantum limit (SQL) is currently within a factor of five away, with the system predominately dominated by thermal noise. By cryogenically cooling the system, this paves the way in reaching the SQL \cite{Braginsky_SQL}, and measuring sub-SQL sensitivity with non-classical states of light.

\section*{Acknowledgements}
This research was supported by the Australian Research Council under the ARC Centre of Excellence for Gravitational Wave Discovery, grand number CE170100004. J.C and T.C are supported by the National Science Foundation grant PHY-1150531 and PHY-1806634. B.S. has been supported by ARC Future Fellowship FT130100329.


---
%
%
%


\bibliographystyle{ieeetr}
\bibliography{biblioRPL}

\begin{thebibliography}{10}

\bibitem{Optomechanics}
M.~Aspelmeyer, T.~J. Kippenberg, and F.~Marquardt, ``Cavity optomechanics,''
  {\em Rev. Mod. Phys.}, vol.~86, pp.~1391--1452, Dec 2014.

\bibitem{Caves_1980}
C.~M. Caves, ``Quantum-mechanical radiation-pressure fluctuations in an
  interferometer,'' {\em Phys. Rev. Lett.}, vol.~45, pp.~75--79, Jul 1980.

\bibitem{Braginsky_book}
V.~B. Braginsky and A.~B. Manukin, {\em Measurement of Weak Forces in Physics
  Experiments}.
\newblock University of Chicago Press, 6 1977.

\bibitem{LIGO}
{\relax J. Aasi \textit{et al}}.~{(LIGO Scientific Collaboration)}, ``{Advanced
  LIGO},'' {\em Classical and Quantum Gravity}, vol.~32, no.~7, p.~074001,
  2015.

\bibitem{VIRGO}
{\relax F. Acernese \textit{et al}}.~{(VIRGO Collaboration)}, ``{Advanced
  Virgo: a second-generation interferometric gravitational wave detector},''
  {\em Classical and Quantum Gravity}, vol.~32, no.~2, p.~024001, 2015.

\bibitem{KAGRA}
{\relax K. Somiya \textit{et al}}.~{(KAGRA Collaboration)}, ``{Detector
  configuration of KAGRA–the Japanese cryogenic gravitational-wave
  detector},'' {\em Classical and Quantum Gravity}, vol.~29, no.~12, p.~124007,
  2012.

\bibitem{Caves1981}
C.~M. Caves, ``Quantum-mechanical noise in an interferometer,'' {\em Phys. Rev.
  D}, vol.~23, pp.~1693--1708, Apr 1981.

\bibitem{LIGO_SQZ}
``Enhanced sensitivity of the ligo gravitational wave detector by using
  squeezed states of light,'' {\em Nature Photonics}, vol.~7, pp.~613--619, Jul
  2013.

\bibitem{GEO_SQZ}
H.~Grote, K.~Danzmann, K.~L. Dooley, R.~Schnabel, J.~Slutsky, and H.~Vahlbruch,
  ``First long-term application of squeezed states of light in a
  gravitational-wave observatory,'' {\em Phys. Rev. Lett.}, vol.~110,
  p.~181101, May 2013.

\bibitem{Kimble}
H.~J. Kimble, Y.~Levin, A.~B. Matsko, K.~S. Thorne, and S.~P. Vyatchanin,
  ``Conversion of conventional gravitational-wave interferometers into quantum
  nondemolition interferometers by modifying their input and/or output
  optics,'' {\em Phys. Rev. D}, vol.~65, p.~022002, Dec 2001.

\bibitem{Yiqiu}
Y.~Ma, H.~Miao, B.~H. Pang, M.~Evans, C.~Zhao, J.~Harms, R.~Schnabel, and
  Y.~Chen, ``Proposal for gravitational-wave detection beyond the standard
  quantum limit through {EPR} entanglement,'' {\em Nature Physics}, vol.~13,
  pp.~776 EP --, 05 2017.

\bibitem{Negative_mass}
F.~Y. Khalili and E.~S. Polzik, ``Overcoming the standard quantum limit in
  gravitational wave detectors using spin systems with a negative effective
  mass,'' {\em Phys. Rev. Lett.}, vol.~121, p.~031101, Jul 2018.

\bibitem{Purdy}
T.~P. Purdy, R.~W. Peterson, and C.~A. Regal, ``Observation of radiation
  pressure shot noise on a macroscopic object,'' {\em Science}, vol.~339,
  no.~6121, pp.~801--804, 2013.

\bibitem{Teufel}
J.~D. Teufel, F.~Lecocq, and R.~W. Simmonds, ``Overwhelming thermomechanical
  motion with microwave radiation pressure shot noise,'' {\em Phys. Rev.
  Lett.}, vol.~116, p.~013602, Jan 2016.

\bibitem{Clark_QRPN_SQZ}
J.~B. Clark, F.~Lecocq, R.~W. Simmonds, J.~Aumentado, and J.~D. Teufel,
  ``Observation of strong radiation pressure forces from squeezed light on a
  mechanical oscillator,'' {\em Nature Physics}, vol.~12, Mar 2016.

\bibitem{Purdy_room_T}
T.~P. Purdy, K.~E. Grutter, K.~Srinivasan, and J.~M. Taylor, ``Quantum
  correlations from a room-temperature optomechanical cavity,'' {\em Science},
  vol.~356, no.~6344, pp.~1265--1268, 2017.

\bibitem{Sudhir}
V.~Sudhir, R.~Schilling, S.~A. Fedorov, H.~Sch\"utz, D.~J. Wilson, and T.~J.
  Kippenberg, ``Quantum correlations of light from a room-temperature
  mechanical oscillator,'' {\em Phys. Rev. X}, vol.~7, p.~031055, Sep 2017.

\bibitem{Cripe_QRPN}
J.~Cripe, N.~Aggarwal, B.~Lanza, A.~Libson, R.~Singh, P.~Heu, D.~Follman, G.~D.
  Cole, N.~Mavalvala, and T.~Corbitt, ``Observation of a room-temperature
  oscillator's motion dominated by quantum fluctuations over a broad
  audio-frequency band,'' 2018.

\bibitem{cole08}
G.~D. Cole, S.~Gr{\"o}blacher, K.~Gugler, S.~Gigan, and M.~Aspelmeyer,
  ``Monocrystalline {A}l$_\mathrm{x}${G}a$_{1 - \mathrm{x}}${A}s
  heterostructures for high-reflectivity high-q micromechanical resonators in
  the megahertz regime,'' {\em Applied Physics Letters}, vol.~92, no.~26,
  p.~261108, 2008.

\bibitem{cole12}
G.~D. Cole, ``Cavity optomechanics with low-noise crystalline mirrors,'' in
  {\em Proc. SPIE 8458, Optics \& Photonics, Optical Trapping and Optical
  Micromanipulation IX}, p.~845807, SPIE, August 2012.

\bibitem{cole13}
G.~D. Cole, W.~Zhang, M.~J. Martin, J.~Ye, and M.~Aspelmeyer, ``Tenfold
  reduction of brownian noise in high-reflectivity optical coatings,'' {\em Nat
  Photon}, vol.~7, pp.~644--650, Aug. 2013.

\bibitem{cole14}
G.~D. Cole, W.~Zhang, B.~J. Bjork, D.~Follman, P.~Heu, C.~Deutsch,
  L.~Sonderhouse, J.~Robinson, C.~Franz, A.~Alexandrovski, M.~Notcutt, O.~H.
  Heckl, J.~Ye, and M.~Aspelmeyer, ``High-performance near- and mid-infrared
  crystalline coatings,'' {\em Optica}, vol.~3, pp.~647--656, Jun 2016.

\bibitem{Singh_PRL}
R.~Singh, G.~D. Cole, J.~Cripe, and T.~Corbitt, ``Stable optical trap from a
  single optical field utilizing birefringence,'' {\em Phys. Rev. Lett.},
  vol.~117, p.~213604, Nov 2016.

\bibitem{13}
B.~S. Sheard, M.~B. Gray, C.~M. Mow-Lowry, D.~E. McClelland, and S.~E.
  Whitcomb, ``Observation and characterization of an optical spring,'' {\em
  Phys. Rev. A}, vol.~69, p.~051801, May 2004.

\bibitem{17}
T.~Corbitt, Y.~Chen, E.~Innerhofer, H.~M\"uller-Ebhardt, D.~Ottaway,
  H.~Rehbein, D.~Sigg, S.~Whitcomb, C.~Wipf, and N.~Mavalvala, ``An all-optical
  trap for a gram-scale mirror,'' {\em Phys. Rev. Lett.}, vol.~98, p.~150802,
  Apr 2007.

\bibitem{Cripe_RPL}
J.~Cripe, N.~Aggarwal, R.~Singh, R.~Lanza, A.~Libson, M.~J. Yap, G.~D. Cole,
  D.~E. McClelland, N.~Mavalvala, and T.~Corbitt, ``Radiation-pressure-mediated
  control of an optomechanical cavity,'' {\em Phys. Rev. A}, vol.~97,
  p.~013827, Jan 2018.

\bibitem{Drever1983}
R.~W.~P. Drever, J.~L. Hall, F.~V. Kowalski, J.~Hough, G.~M. Ford, A.~J.
  Munley, and H.~Ward, ``Laser phase and frequency stabilization using an
  optical resonator,'' {\em Applied Physics B}, vol.~31, no.~2, pp.~97--105,
  1983.

\bibitem{Chua11}
S.~S.~Y. Chua, M.~S. Stefszky, C.~M. Mow-Lowry, B.~C. Buchler, S.~Dwyer, D.~A.
  Shaddock, P.~K. Lam, and D.~E. McClelland, ``Backscatter tolerant squeezed
  light source for advanced gravitational-wave detectors,'' {\em Opt. Lett.},
  vol.~36, pp.~4680--4682, Dec 2011.

\bibitem{Henning}
H.~Vahlbruch, S.~Chelkowski, B.~Hage, A.~Franzen, K.~Danzmann, and R.~Schnabel,
  ``Coherent control of vacuum squeezing in the gravitational-wave detection
  band,'' {\em Phys. Rev. Lett.}, vol.~97, p.~011101, Jul 2006.

\bibitem{Braginsky547}
V.~B. Braginsky, Y.~I. Vorontsov, and K.~S. Thorne, ``Quantum nondemolition
  measurements,'' {\em Science}, vol.~209, no.~4456, pp.~547--557, 1980.

\bibitem{Braginsky_speedmeter}
V.~B. Braginsky, M.~L. Gorodetsky, F.~Y. Khalili, and K.~S. Thorne,
  ``Dual-resonator speed meter for a free test mass,'' {\em Phys. Rev. D},
  vol.~61, p.~044002, Jan 2000.

\bibitem{Harms}
J.~Harms, Y.~Chen, S.~Chelkowski, A.~Franzen, H.~Vahlbruch, K.~Danzmann, and
  R.~Schnabel, ``Squeezed-input, optical-spring, signal-recycled
  gravitational-wave detectors,'' {\em Phys. Rev. D}, vol.~68, p.~042001, Aug
  2003.

\bibitem{FD_SQZ}
E.~Oelker, T.~Isogai, J.~Miller, M.~Tse, L.~Barsotti, N.~Mavalvala, and
  M.~Evans, ``Audio-band frequency-dependent squeezing for gravitational-wave
  detectors,'' {\em Phys. Rev. Lett.}, vol.~116, p.~041102, Jan 2016.

\bibitem{Glasgow_speedmeter}
\relax C.~Gräf~\textit{et al}., ``Design of a speed meter interferometer
  proof-of-principle experiment,'' {\em Class. Quantum Grav.}, vol.~31,
  p.~215009, 2014.

\bibitem{Giovannetti1330}
V.~Giovannetti, S.~Lloyd, and L.~Maccone, ``Quantum-enhanced measurements:
  Beating the standard quantum limit,'' {\em Science}, vol.~306, no.~5700,
  pp.~1330--1336, 2004.

\bibitem{Braginsky_SQL}
V.~B. Braginsky, ``Classical and quantum restrictions on the detection of weak
  disturbances of a macroscopic oscillator,'' {\em Sov. J. Exp. Theor. Phys.},
  vol.~26, p.~831, 1968.

\end{thebibliography}

\end{document}